\documentclass[aps,prc,preprint,showpacs,groupedaddress]{revtex4}

\usepackage{graphicx} 
\usepackage{amsmath,amssymb,amsfonts} 
\usepackage{dcolumn} 
\usepackage{bm} 
\usepackage[english]{babel} 

\newcommand{\ssst}{\scriptscriptstyle}
\newcommand{\sst}{\scriptstyle}
\newcommand{\dst}{\displaystyle}
\newcommand{\abs}[1]{\ensuremath{\vert #1 \vert}}

\begin{document}

\title{Forward-angle $\bm{K^+ \Lambda}$ photoproduction in a Regge-plus-resonance approach}

\author{T. Corthals}
\email[]{tamara.corthals@ugent.be}
\author{J. Ryckebusch}
\author{T. Van Cauteren}
\affiliation{Department of Subatomic and Radiation Physics, \\
  Ghent University, Proeftuinstraat 86, B-9000 Gent, Belgium}
\date{\today}

\begin{abstract}
We present an effective-Lagrangian description for forward-angle $K^+\Lambda$ photoproduction from the proton, valid for photon lab energies from threshold up to $16~\mathrm{GeV}$. The high-energy part of the amplitude is modeled in terms of $t$-channel Regge-trajectory exchange. The sensitivity of the calculated observables to the Regge-trajectory phase is investigated in detail. The model is extended towards the resonance region by adding a number of $s$-channel resonances to the $t$-channel background. The proposed hybrid ``Regge-plus-resonance'' (RPR) approach allows one to exploit the $p(\gamma,K^+)\Lambda$ data in their entirety, resulting in stronger constraints on both the background and resonance couplings. The high-energy data can be used to fix the background contributions, leaving the resonance couplings as the sole free parameters in the resonance region. We compare various implementations of the RPR model, and explore to what extent the description of the data can be improved by introducing the ``new'' resonances $D_{13}(1900)$ and $P_{11}(1900)$. Despite its limited number of free parameters, the proposed RPR approach provides an efficient description of the $p(\gamma,K^+)\Lambda$ dynamics in and beyond the resonance region.
\end{abstract}


\maketitle

\section{Introduction}

The study of photo- and electroinduced associated open-strangeness production from the nucleon is playing an increasingly important role in unraveling the structure of matter at hadronic energy scales. Recent measurements performed at the JLab, ELSA and SPring-8 facilities have resulted in an impressive set of precise $p(\gamma,K^+)\Lambda$ data in the few-GeV regime \cite{McNabb03,Brad05,Glander04,Zegers03,Sumihama05}, comprising differential cross sections as well as hyperon polarizations and photon beam asymmetries. These experimental achievements have motivated renewed efforts by various theoretical groups \cite{MaSuBe04,MoShklyar05,Diaz05,Sara05}. A great deal of attention has been directed towards the development of tree-level isobar models~\cite{Stijnprc01,AdWr88,WJ92,DaSa96,MaBeHy95,MaBe00}, in which the scattering amplitude is constructed from a number of lowest-order Feynman diagrams. In such a framework, the challenge lies in selecting the relevant set of $s$-channel resonance diagrams to be added to the background. The latter consists of the Born terms, complemented by a number of $t$- and/or $u$-channel single-particle exchange diagrams. While providing a satisfactory description of photo- and electroproduction observables for energies up to 2-3 GeV, such a lowest-order approximation has its limitations. Since decay widths must be introduced to account for the resonances' finite lifetimes, the unitarity requirement is not fulfilled. Further, higher-order mechanisms like final-state interactions and channel couplings are not explicitly included. Chiang \emph{et al.~}\cite{Chiang01} have shown that the contribution of the intermediate $\pi N$ channel to the $p(\gamma,K^+)\Lambda$ cross sections is of the order of 20\%. Sever\-al groups are presently engaged in efforts to extract the resonance parameters in a full-blown coupled-channels analysis~\cite{Mosel02_pho,MoShklyar05,Diaz05,US05}. 

Admittedly, it is debatable whether an extraction of resonance information can be reliably performed at tree level. It should be stressed, however, that the current description of some channels, including the $K\Lambda$ one, is plagued by severe uncertainties. The choice of gauge restoration procedure~\cite{gauge_inv}, for example, has recently been demonstrated to have a large impact on the computed observables \cite{US05}. Also, a fundamental understanding of the functional form of the hadronic form factors and the magnitude of the cutoff values is still lacking. We deem that these issues can better be addressed at the level of the individual reaction channels, where the number of parameters and uncertainties can be kept at a manageable level. 

A major challenge for tree-level descriptions of electromagnetic $KY$ production is modeling the background contribution to the amplitude. Several background models have been proposed~\cite{WJ92,MaBe00,Stijnprc01}, differing primarily in the mechanism used for reducing the Born-term contribution, which in itself spectacularly overshoots the measured cross sections. Though all of the proposed strategies allow for a reasonable description of the data, the extracted resonance couplings are quite sensitive to the specifics of the background model~\cite{GApaper03}. 

In this paper, we shall explore an alternative tactic for constraining the $p(\gamma,K^+)\Lambda$ background dynamics. The procedure is based on fixing the background couplings to the high-energy data. At energies $\omega_{lab} \gtrsim 4$ GeV, the forward-angle behavior of the observables can be described in a model based on Regge-pole exchange in the $t$ channel \cite{reg_guidal, reg_guidal_2,Levy73}. Designed as a high-energy phenomenological tool, the Regge formalism cannot \emph{a priori} be expected to remain valid in the resonance region. However, it has been found that even meson production data at center-of-mass energies below 2 GeV are reproduced quite well in a Regge model. This is not only the case for the pseudoscalar $K$ and $\pi$ mesons~\cite{Guidalthes}, but also for vector particles like the $\omega$~\cite{Sibi03}. Accordingly, we assume that the Regge parameterization for the high-energy $p(\gamma,K^+)\Lambda$ background remains physical in the resonance region. Evidently, the additional structures appearing in the cross sections at lower energies cannot be reproduced in a pure background model. This can be remedied by superimposing a number of $s$-channel resonances onto the $t$-channel background. A similar strategy was applied to high-energy double-pion production in Ref.~\cite{Holvoet_phd}, and to the production of $\eta$ and $\eta'$ mesons in Ref.~\cite{Chi03}. In this paper, a ``Regge-plus-resonance'' (RPR) model for the $p(\gamma,K^+)\Lambda$ process is developed.

The RPR approach has a number of assets. Firstly, the correct high-energy behavior of the observables is automatically ensured, whereas isobar models based on single-particle exchange are destined to fail in the high-energy limit. Secondly, the background dynamics can be distilled from the high-energy data, leaving the resonance couplings as the sole parameters to be determined in the resonance region. 

This paper is organized as follows. We outline the different aspects of the RPR model in Sec.~\ref{sec: RPR}. Sec.~\ref{subsec: RPR background} focuses on the procedure for $t$-channel reggeization of the high-energy amplitude. In Sec.~\ref{subsec: RPR res}, we suggest an extension of the Regge model aimed at incorporating resonance dynamics. Our numerical results for the $p(\gamma,K^+)\Lambda$ process, covering forward angles and photon lab energies from threshold up to 16 GeV, are presented in Sec.~\ref{sec:results}. In Sec.~\ref{subsec: results background} it is shown that the background dynamics can be well constrained by the high-energy observables, in particular by the recoil polarization. Sec.~\ref{subsec: results res} summarizes the results of our calculations in the resonance region. Finally, in Sec.~\ref{sec: conclusion} we state our conclusions.

\section{RPR model for forward-angle $\bm{K^+\Lambda}$ production}
\label{sec: RPR}

\subsection{Regge description of high-energy dynamics}
\label{subsec: RPR background}

The strength of the formalism introduced by Regge in 1959 \cite{Regge59} lies primarily in its elegant treatment of high-spin, high-mass particle exchange. Rather than focusing on individual particles, Regge theory considers the exchange of entire \emph{families} of hadrons, with identical internal quantum numbers but different spins $J$. The members of a family are connected by an approximately linear relation between their spin and squared mass, and are said to lie on a \emph{Regge trajec\-tory}. The Regge formalism has been specifically designed for processes at high $s$ and at small $\abs{t}$ or $\abs{u}$, corresponding to forward and backward scattering angles respectively. In this regime, the dynamics is governed by the exchange of one or more Regge trajectories. At forward angles this exchange takes place in the $t$ channel, at backward angles in the $u$ channel. The resulting amplitude is proportional to $s$ raised to a negative power, and thus obeys the Froissart bound \cite{Froiss_61}, which is a necessary condition for unitarity. It is worth noting that the Froissart bound is violated in an isobar framework, where the exchange of a nonscalar particle leads to an amplitude increasing with energy faster than $\log^2 s$.

Here, we focus on the forward-angle $p(\gamma,K^+)\Lambda$ dynamics, modeled by meson trajectory exchange in the $t$ channel. Each trajectory is characterized by a linear function $\alpha(t)$ of the exchanged four-momentum squared, with the spins and masses of the physical particles obeying the relation ${J=\alpha\,({t=m^2})}$. The functions $\alpha(t)$ correspond to poles of the scattering amplitude in the complex angular momentum plane.

\begin{figure}[b]
\begin{center}
\includegraphics[width=0.58\textwidth, clip]{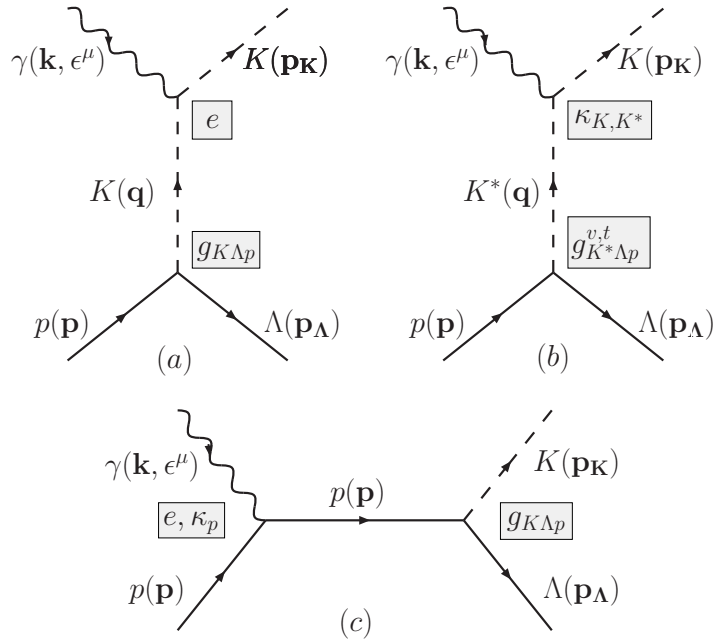}
\caption{Feynman graphs contributing to the $p(\gamma,K^+)\Lambda$ amplitude for  $\omega_{lab}\gtrsim 4$ GeV and at forward angles: exchange of (a) $K$ and (b) $K^{\ast}$ trajectories. The electric part of the $s$-channel Born term [diagram (c)] is added to restore gauge invariance.}
\label{fig:feyndiag_t}
\end{center}
\end{figure}

\begin{figure}[t]
\begin{center}
\includegraphics[width=0.47\textwidth]{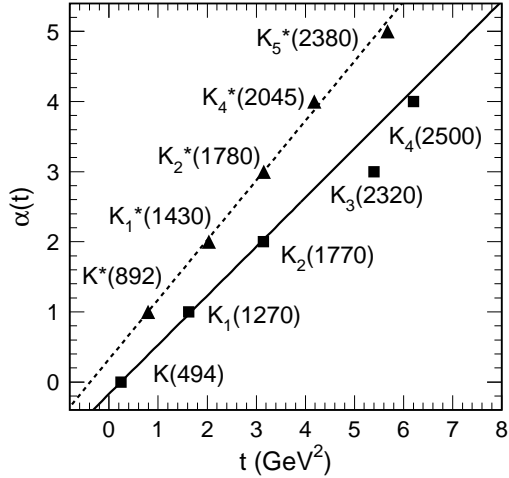}
\caption{Chew-Frautschi plots for the meson trajectories in the kaon sector. The meson masses are from the Particle Data Group~\cite{PDG04}.}
\label{fig: kaon-traj}
\end{center}
\end{figure}

As we aim at developing a consistent description of the $p(\gamma,K^+)\Lambda$ observables in and above the resonance region, we embed the Regge formalism into an effective-field model. At tree level, the relevant parameters are simply products of a strong and an electromagnetic coupling. When reggeizing the $p(\gamma,K^+)\Lambda$ scattering amplitude, only the Feynman diagrams for the lowest-lying members, called \emph{first materializations}, of the various meson trajectories $\alpha(t)$ are retained. In the corresponding amplitudes, the Feynman propagators are replaced by Regge propagators through the substitution
\begin{equation}
\frac{1}{t-m^2} \hspace{6pt} \xrightarrow{~\quad} \hspace{8pt} \mathcal{P}_{Regge}\bigl(\alpha(t)\bigr)\,.
\end{equation}
The diagrams contributing to the high-energy, forward-angle $K^+\Lambda$ photoproduction amplitude are shown in Fig.~\ref{fig:feyndiag_t}. We will refer to them as background terms, since none of them passes through a pole in the physical plane of the $p(\gamma,K^+)\Lambda$ process.
The Feynman graphs at the top of Fig.~\ref{fig:feyndiag_t} represent the $t$-channel exchanges of the $K(494)$ (a) and $K^{\ast}(892)$ (b) trajectories. The corresponding $J$ versus $m^2$ plots (Chew-Frautschi plots) for the trajectory members are shown in Fig.~\ref{fig: kaon-traj}. These two diagrams would suffice to describe the high-energy amplitude, were it not for the $K$-exchange contribution which breaks gauge invariance. It is a property inherent to the effective-Lagrangian framework that the individual Born terms in the $s$, $t$ and $u$ channel do not obey gauge invariance, while their sum does. The $p(\gamma,K^+)\Lambda$ amplitude can therefore be made gauge invariant by adding the \emph{electric} part of the $s$-channel Born diagram~[Fig.~\ref{fig:feyndiag_t}(c)] according to the recipe~\cite{reg_guidal, reg_guidal_2}
\begin{equation}
\begin{split}
\mathcal{M}\,(&\gamma\,p \rightarrow K^+\Lambda) =~ \mathcal{M}_{Regge}^K + \mathcal{M}_{Regge}^{K^{\ast}} \quad \\ 
& + \mathcal{M}_{Feyn}^{p\ssst,\sst elec} \times \  \mathcal{P}_{Regge}^K \times \ (t-m_K^2)\,.
\label{eq: gauge_recipe}
\end{split}
\end{equation}
In Sec.~\ref{sec:results}, we will show that implementing this gauge-invariance res\-toration procedure leads to an improved description of the $p(\gamma,K^+)\Lambda$ differential cross section at $\vert t \vert \rightarrow 0$. 

A derivation of the Regge amplitude for spinless external particles can \emph{e.g.}~be found in Ref.~\cite{Donnachie02}. In the high-$s$, low-$\vert t \vert$ limit, the result can be written as
\begin{equation}
\begin{split}
\mathcal{M}^{\ \zeta=\pm}_{Regge}(s,t) \ =&  \quad C  \, \left(\frac{\dst s}{\dst s_0}\right)^{\alpha_{\zeta}(t)} \frac{\beta(t)}{\sin\bigl(\pi\alpha_{\zeta}(t)\bigr)} \\ 
& \quad \times ~ \frac{1 + \zeta\,e^{-i\pi\alpha_{\zeta}(t)}}{2}  \ \frac{1}{\Gamma\bigl(1 + \alpha_{\zeta}(t)\bigr)}\,,
\label{eq: spinlessregge}
\end{split}
\end{equation}
with the scale factor $s_0$ fixed at $1~\mathrm{GeV}^2$. In deriving this result, one has to distinguish between the two signature parts $\alpha_+(t)$ and $\alpha_-(t)$ of the trajectory in order to satisfy the convergence criteria. The above amplitude has poles at values of $t$ where $\alpha_{\zeta}(t)$ assumes a non-negative even ($\zeta=+$) or odd ($\zeta=-$) integer value. These values are precisely the spins of the particles lying on the $\alpha_{\pm}(t)$ Regge trajectories, with the corresponding values of $t$ equaling the particles' squared masses. Thus, the $\zeta=+$ trajectory connects particles with $J= 0, 2, 4$ etc.,~and is of the form 
\begin{equation}
\alpha_+(t) = \alpha_+' (t-m_0^2)\,,
\end{equation}
with $m_0$ the mass of the spin-0 first materialization. Similarly, the negative-signature trajectory can be written as $\alpha_-(t)=1 + \alpha_- ' (t-m_1^2)$.~Unknowns in Eq.~(\ref{eq: spinlessregge}) are the constant $C$ and the residue function $\beta(t)$. They can be determined by linking the $\gamma\,p \rightarrow K^+\Lambda$ amplitude to the amplitude of the crossed $t$-channel process $\gamma K^- \rightarrow \bar{p} \Lambda$. Crossing symmetry implies that both processes can be described by the same function $\mathcal{M}(s,t)$ in the complex $(s,t)$ plane, albeit with the two Mandelstam variables $s$ and $t$ interchanged. Regge phenomenology exploits this symmetry by analytically continuing the reaction amplitude from the $t$-channel physical region into the $s$-channel physical region. In the vicinity of a $t$-channel pole $m_X^2$, the amplitude for the crossed process reduces to
\begin{equation}
\mathcal{M}^{\,\gamma K^- \rightarrow \bar{p} \Lambda}_{Feyn}(t,s) \quad \xrightarrow{t=m_X^2} \ \frac{\beta_X(t)}{t-m_X^2}\,.
\end{equation}
We now demand that the crossed amplitude, when evaluated at its $t$-channel pole closest to the $\gamma\,p \rightarrow K^+\Lambda$ physical region (where $t<0$), equals the Regge amplitude (\ref{eq: spinlessregge}). Thus, for the $\zeta=+$ case, we have the requirement
\begin{equation}
\begin{split}
\mathcal{M}_{Regge}^{\,\gamma p \rightarrow K^+ \Lambda,~\zeta=+}&(s,t) \quad =^{\substack{\hspace{-4mm}\vspace{2mm} t=m_0^2}} \\ & \mathcal{M}^{\,\gamma K^- \rightarrow \bar{p} \Lambda}_{~Feyn}(t,s) = \frac{\beta_0(t)}{t-m_0^2}\,.
\label{feyn_reg_eq}
\end{split}
\end{equation}
If $\beta(t)$ is taken to be equal to the residue $\beta_0(t)$ of the crossed Feynman amplitude, Eq.~(\ref{feyn_reg_eq}) leads to $C=\pi\alpha'_+$. If we now define
\begin{equation}
\mathcal{M}_{Regge}(s,t) = \mathcal{P}_{Regge}(s,t) \times \beta(t)\,,
\label{eq: propdef}
\end{equation}
we obtain for the Regge propagator:
\begin{equation}
\begin{split}
\mathcal{P}^{\ \zeta=\pm}_{Regge} = \frac{ \left(\frac{\dst s}{\dst s_0}\right)^{\alpha(t)}}{\sin\bigl(\pi\alpha(t)\bigr)} \ \frac{1 + \zeta\,e^{-i\pi\alpha(t)}}{2}  \ \frac{\pi\alpha'}{\Gamma\bigl(1 + \alpha(t)\bigr)}\,.
\end{split}
\label{eq: reggeprop_spinless}
\end{equation}

Often, the positive- and negative-signature parts of a trajectory coincide.~If, in addition, the corresponding residues $\beta(t)$ are identical or differ only in their sign, the trajectory parts are called \emph{strongly degenerate}. Then, the $\zeta=\pm$ amplitudes differ only by their phase, and in determining the total amplitude the propagators $\mathcal{P}^{\ \zeta=\pm}_{Regge}$ can either be added or subtracted. As a consequence, the Regge propagator for a degenerate trajectory can have a constant or a rotating phase: 
\begin{equation}
\mathcal{P}_{Regge} = \frac{ \left(\frac{\dst s}{\dst s_0}\right)^{\alpha(t)}}{\sin\bigl(\pi\alpha(t)\bigr)} \
\left\{ \begin{array}{c}
1 \\
e^{-i\pi\alpha(t)}
\end{array}\right\} \
\frac{\pi\alpha'}{\Gamma\bigl(1 + \alpha(t)\bigr)}\,.
\label{eq: reggeprop_spinless_degen}
\end{equation}

Whether or not a trajectory should be treated as degenerate depends on the process under study. Non-degenerate trajectories give rise to dips in the differential cross section because they exhibit so-called \emph{wrong-signature zeroes}~\cite{Collins77}. These are zeroes of the Regge propagator corresponding to poles of the gamma function which are not removed by the phase factor. \emph{E.g.}, for $\zeta=+$ the propagator (\ref{eq: reggeprop_spinless}) has wrong-signature zeroes at strictly negative, odd values of $\alpha(t)$. Vice versa, a smooth, structureless cross-section points to degenerate trajectories. Since no obvious structure is present in the $p(\gamma,K^+)\Lambda$ cross-section data for $\omega_{lab} \gtrsim 4~\mathrm{GeV}$, both the $K$ and $K^{\ast}$ trajectories are assumed to be degenerate. As is clear from Fig. \ref{fig: kaon-traj}, this is certainly justified in the $K^{\ast}$ case. The positive- and negative-signature parts of the $K$ trajectory are not as perfectly collinear.

Generalizing the above results to nonscalar particles is a nontrivial task~\cite{Collins77}. In order to restrict the number of unknown parameters in the $p(\gamma,K^+)\Lambda$ background model, we opt for a more phenomenological approach. A general Regge trajectory is of the form
\begin{equation}
\alpha(t)=\alpha_0 + \alpha'(t-m_0^2)\,,
\end{equation}
with $\alpha_0$ the spin and $m_0$ the mass of the trajectory's first materialization. By replacing $\alpha(t)$ in Eq.~(\ref{eq: reggeprop_spinless_degen}) by $\alpha(t)-\alpha_0$ in the exponent of $s$ and in the argument of the gamma function, it is guaranteed that the condition (\ref{feyn_reg_eq}) is also fulfilled for trajectories with a nonscalar first materialization. The altered gamma function further ensures that the resulting propagator has the correct pole structure, with poles at integer $\alpha(t) \ge \alpha_0$. This results in the following expressions for the $K$ and $K^{\ast}(892)$ Regge propagators:
\begin{align}
&\mathcal{P}^K_{Feyn}(s,t) = \frac{1}{t-m_K^2} \quad \longrightarrow \quad \mathcal{P}^K_{Regge}(s,t) = \nonumber \\
&\qquad \left(\frac{\dst s}{\dst s_0}\right)^{\alpha_K(t)}
\frac{1}{\sin\bigl(\pi\alpha_K(t)\bigr)}  \hspace{10pt} 
\left\{ \begin{array}{c}
1 \\
e^{-i\pi\alpha_{K}(t)}
\end{array}\right\} \hspace{10pt} 
\frac{\pi \alpha'_K}{\Gamma\bigl(1+\alpha_K(t)\bigr)}\,, \label{eq: reggeprop_K}
\end{align}
\begin{align}
&\mathcal{P}^{K^{\ast}}_{Feyn}(s,t) = \frac{1}{t-{m_{K^{\ast}}^2}} \quad \longrightarrow \quad \mathcal{P}^{K^{\ast}}_{Regge}(s,t) = \nonumber \\
&\qquad \left(\frac{\dst s}{\dst s_0}\right)^{\alpha_{K^{\ast}}(t)-1} 
\frac{1}{\sin\bigl(\pi\alpha_{K^{\ast}}(t)\bigr)} \hspace{10pt}
\left\{ \begin{array}{c}
1 \\
e^{-i\pi\alpha_{K^*}(t)}
\end{array}\right\} \hspace{10pt}
\frac{\pi \alpha'_{K^{\ast}}}{\Gamma\bigl(\alpha_{K^{\ast}}(t)\bigr)}\,.\label{eq: reggeprop_Kstar}
\end{align}
The $K$ and $K^{\ast}$ trajectories are given by~\cite{Stijnthes} 
\begin{align}
\alpha_K(t) &= 0.70 \ \mathrm{GeV}^{-2} \ (t-m_K^2)\,,\label{eq: Ktraj}\\
\alpha_{K^{\ast}}(t) &= 1 + 0.85 \ \mathrm{GeV}^{-2} \ (t-m_{K^{\ast}}^2)\,.\label{eq: Kstartraj} 
\end{align}
The total Regge amplitude is con\-struc\-ted by substituting the expressions for the Regge propagators into Eq.~(\ref{eq: gauge_recipe}). In the Appendix we summarize the strong and electromagnetic interaction Lagrangians needed for calculating the Feynman residues $\beta(t)$ in Eq.~(\ref{eq: propdef}). Using these interactions, the background model, which consists solely of $t$-channel diagrams, contains only three parameters:
\begin{equation}
g_{K\Lambda p}\,, \quad G_{K^{\ast}}^{v,t} = \frac{e g_{{K^{\ast}} \Lambda p}^{v,t}}{ 4 \pi} \ \kappa_{K{K^{\ast}}}\;.
\label{eq: bg_free_pars}
\end{equation}
These parameters will be fitted against the high-energy observables. 

We have deliberately chosen not to treat $u$-channel reggeization at this time. One reason for this is the scarcity of high-energy data in the backward-angle regime. The second, more fundamental reason involves the fact that the lightest hyperon, the $\Lambda$, is significantly heavier than a $K$ meson. As a consequence, the $u$-channel poles are much further removed from the backward-angle kinematical regime than the $t$-channel poles are from the forward-angle region. Accordingly, for $u$-channel reggeization, the procedure of requiring the Regge propagator to reduce to the Feynman one at the closest crossed-channel pole cannot be guaranteed to lead to good results.

\subsection{Inclusion of resonance dynamics}
\label{subsec: RPR res}

Regge theory is essentially a high-energy approach. Accordingly, the Regge amplitudes based on the propagators of Eqs.~(\ref{eq: reggeprop_K}) and (\ref{eq: reggeprop_Kstar}) should be interpreted as the asymptotic forms of the full amplitudes for $s \rightarrow \infty$, $\abs{t}\rightarrow 0$. The experimental meson production cross sections appear to exhibit this ``asymptotic'' Regge behavior for photon energies down to about 4 GeV~\cite{reg_guidal, reg_guidal_2,Sibi03}. In the resonance region, on the other hand, a Regge-pole description can no longer be expected to account for all aspects of the reaction dynamics. At low energies, the cross sections exhibit more pronounced structures, such as peaks at certain energies and complex variations in the angular distributions. There exists, however, a theoretical connection between the high- and low-energy domain, which is related to the notion of \emph{duality}. Simply put, the duality hypothesis states that, on average, the sum of all resonant contributions in the $s$ channel equals the sum of all Regge poles exchanged in the $t$ channel. In practice, it is impossible to take all $s$-channel diagrams into account. Hence, the standard procedure consists of identifying a small number of dominant resonances, and supplementing these with a phenomenological background. 

We adopt the Regge description for both the high-energy amplitude and the background contribution to the resonance-region amplitude. Indeed, in Ref.~\cite{Guidalthes} it is demonstrated that, even when using the asymptotic form of the propagators, the order of magnitude of the forward-angle pion and kaon photoproduction observables in the resonance region is remarkably well reproduced in a pure $t$-channel Regge model. A similar observation holds for kaon electroproduction~\cite{Guidal_elec}. These results imply that at forward angles, the global features of the $p(\gamma,K^+)\Lambda$ reaction in the resonance region can be reasonably well reproduced in terms of background diagrams, with resonance effects constituting relatively minor corrections. This observation suggests that the double-counting issue which arises from superimposing a (small) number of resonances onto the Regge background, may not be a severe one.

A great asset of the RPR approach lies in its elegant description of the non-resonant part of the reaction amplitude. In standard isobar approaches, the determination of the background requires a significantly larger number of parameters. A typical isobar background amplitude consists of Born terms ($p$, $K$, $\Lambda$ and $\Sigma^0$ exchange) complemented by $K^{\ast}(892)$ and $K_1(1270)$ exchange diagrams. Thus, in an isobar model, at least three additional coupling constants, $g_{K\Sigma^0 p}$ and $G_{K_1}^{v,t}$, enter the problem. In some cases, $u$-channel hyperon resonances are introduced. A Regge-inspired model, on the other hand, contains either $t$- or $u$-channel exchanges. In addition, the issue of unreasonably large Born-term strength, which constitutes a major challenge for isobar models, does not arise in the RPR approach. Consequently, no strong form factors are required for the background terms and the introduction of an additional background cutoff parameter is avoided. The Regge model faces only one uncertainty, namely the choice between constant or rotating $K$ and $K^{\ast}$ trajectories.

In the RPR framework, the resonant $s$-channel terms are described using Feynman propagators. The resonances' finite lifetimes are taken into account through the substitution
\begin{equation}
s - m_{N^{\ast}}^2 \longrightarrow ~ s - m_{N^{\ast}}^2 + im_{N^{\ast}}\Gamma_{N^{\ast}}
\end{equation}
in the propagator denominators. The strong and electromagnetic interaction Lagrangians for coupling to spin-1/2 and spin-3/2 resonances are given in the Appendix. 

\begin{figure}[ht] 
\begin{center}
\includegraphics[width=0.63\textwidth]{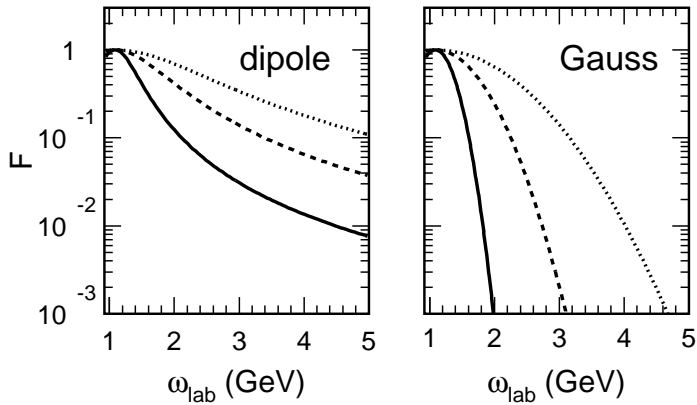}
\caption{Dipole and Gaussian form factors as a function of the photon lab energy $\omega_{lab}$, for a resonance with mass $m_{N^{\ast}}=1710~\mathrm{MeV}$. The full, dashed and dotted curves correspond to cutoffs $\Lambda_{res} = 800$, 1200 and 1600 $\mathrm{MeV}$ respectively.}
\label{fig: strongff}
\end{center}
\end{figure}

In conventional isobar models, the resonance contributions increase with energy. For our RPR approach to be meaningful, however, the resonance amplitudes should vanish at high values of $\omega_{lab}$. This is accomplished by including a phenomenological form factor $F(s)$ at the strong $K\Lambda N^{\ast}$ vertices. Instead of the standard dipole parameterization
\begin{equation}
F_{dipole}(s) = \frac{\Lambda_{res}^4}{\Lambda_{res}^4 + (s-m^2_{N^{\ast}})^2}\,,
\label{eq: dipoleff}
\end{equation}
used in most isobar models, we assume a Gaussian shape
\begin{equation} 
F_{Gauss}(s) = \exp \left\{- \frac{(s-m^2_{N^{\ast}})^2}{\Lambda_{res}^4}\right\}\,,
\label{eq: gaussff}
\end{equation}
with $\Lambda_{res}$ the cutoff value. Both forms are compared in Fig.~\ref{fig: strongff}. 
Our primary motivation for introducing Gaussian form factors is that they fall off much more sharply with energy than dipoles. Using Gaussian form factors, for $\omega_{lab} \gtrsim 4~\mathrm{GeV}$ the resonant contributions to the observables are quenched almost completely, even for cutoff values of $1600~\mathrm{MeV}$ and larger. A comparable effect is impossible to attain with a dipole, even with a cutoff mass as small as $800~\mathrm{MeV}$.

Previous analyses identified the $S_{11}(1650)$, $P_{11}(1710)$ and $P_{13}(1720)$ states as the main resonance contributions to the $p(\gamma,K^+)\Lambda$ dynamics~\cite{FeMo99,MaBe00,Saghai01}. Very recently the importance of the $P_{11}(1710)$ state has been called into question~\cite{Sara05, MoShklyar05}. A new $D_{13}$ resonance was first introduced in Ref.~\cite{MaBe00} to explain a structure in the old SAPHIR total cross-section data~\cite{Tran98} at $W \approx 1900~\mathrm{MeV}$. Hitherto unobserved in $\pi N$ reactions, this $D_{13}$ state has been predicted in constituent-quark model calculations of Capstick and Roberts with a significant branching into the $K\Lambda$ channel~\cite{CapRob94}.~The evidence for this ``missing'' resonance is far from conclusive, however.~Results of more recent analyses \cite{Saghai01,MaSuBe04,Sara05} seem to contradict the previously drawn conclusions. Ref.~\cite{GApaper04} specifically points to a $P_{11}(1900)$ state as a more likely missing-resonance candidate, while the results from Ref.~\cite{Diaz05} suggest that a third $S_{11}$ resonance might be playing a role. A recent coupled-channels study of the $(\pi,\gamma)N\rightarrow K \Lambda$ reactions revealed no evidence for any missing resonance; inclusion of the known $P_{13}(1900)$ state proved sufficient to describe the measured structures in the $p(\gamma,K^+)\Lambda$ observables~\cite{MoShklyar05}. 

Several possible interpretations for the $W \approx 1900~\mathrm{MeV}$ cross-section peak will be explored in this work. In Sec.~\ref{subsec: results res}, we shall present results of numerical calculations performed with different combinations of known and (as yet) unobserved resonances. 

\section{Results and discussion}
\label{sec:results}

\subsection{High-energy data}
\label{subsec: results background}

In contrast to the favorable experimental situation in the resonance region, the available data for $\omega_{lab} \gtrsim 4~\mathrm{GeV}$ are scanty, and mostly correspond to forward-angle kinematics. The relevant low-$\abs{t}$ data comprise 56 differential cross sections in total, at the selected energies $\omega_{lab}=5$, 8, 11 and $16~\mathrm{GeV}$~\cite{Boyarski69}. A limited number of polarization observables are available, in the form of 7 recoil and 9 photon beam asymmetry points at $\omega_{lab} = 5$ and $16~\mathrm{GeV}$ respectively~\cite{Qui79,Vo72}. 

For the high-energy observables, we rely on the pure $t$-channel Regge framework as outlined in Sec.~\ref{subsec: RPR background}. Contrary to the nonstrange sector, where the strong couplings can be determined from $NN$ and $\pi N$ scattering, the hyperon-nucleon interaction is difficult to access. Accordingly, the model parameters defined in Eq.~(\ref{eq: bg_free_pars}) have to be optimized against the aforementioned high-energy data. Assuming $SU(3)$-flavor symmetry, the strong $g_{K\Lambda p}$ coupling constant can be related to the well-known $g_{\pi NN}$ coupling~\cite{deSwart,Mac87}. Assuming that the $SU(3)$ prediction can deviate up to 20\% from the physical value, the following range emerges:
\begin{equation}
-4.5 ~ \le ~ \frac{g_{K\Lambda p}}{\sqrt{4\pi}} ~ \le ~ -3.0\,.
\end{equation}

\begin{figure}[t]
\begin{center}
\includegraphics[width=0.7\textwidth]{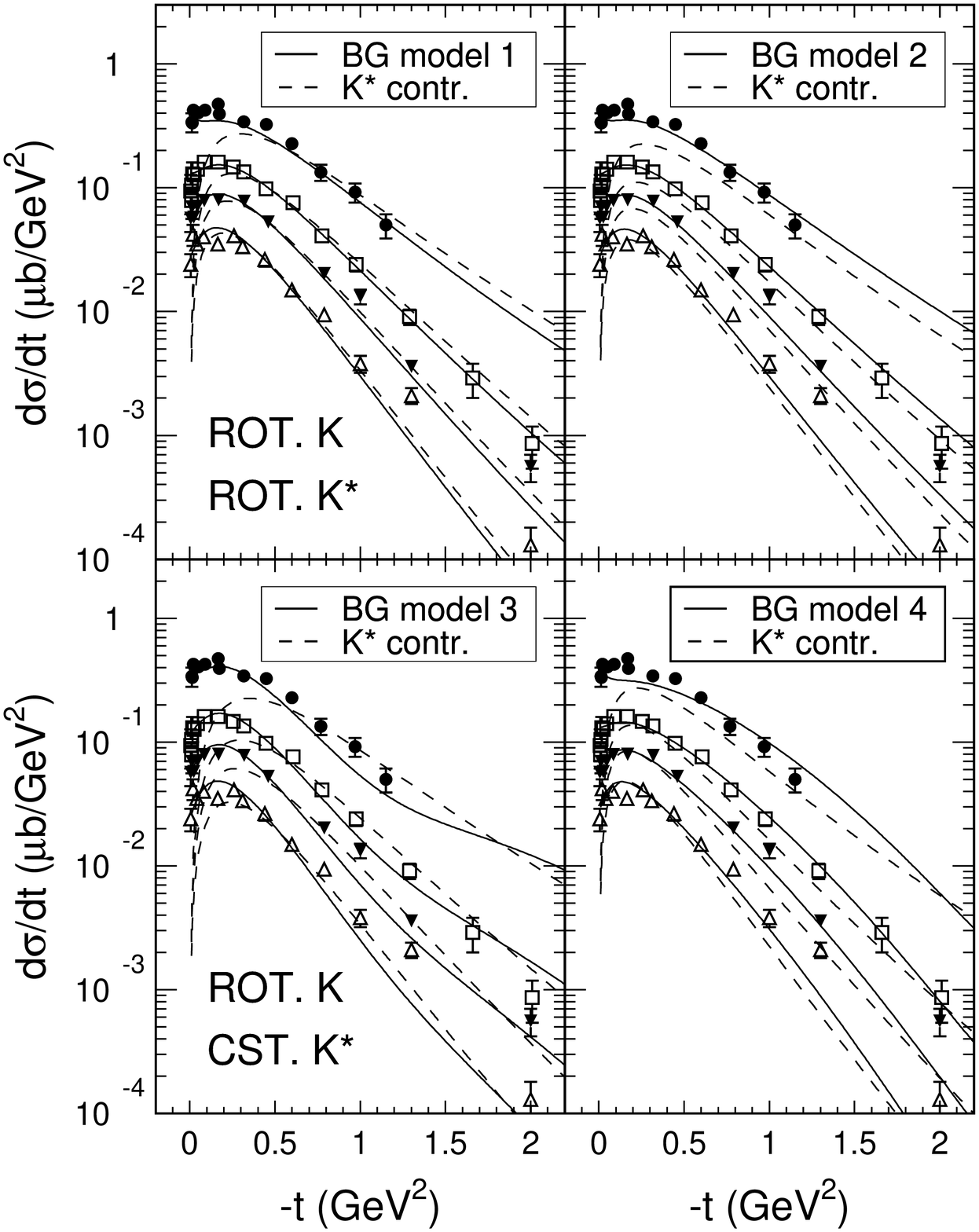}
\caption{Forward-angle differential $p(\gamma,K^+)\Lambda$ cross sections at photon lab energies of 5~($\bullet$), 8~($\square$), 11~($\blacktriangledown$) and 16~($\bigtriangleup$) $\mathrm{GeV}$. The upper panels correspond to the Regge model variants 1 and 2, with a rotating phase for the $K$ and $K^{\ast}$ trajectories. In the lower panels, model variants 3 and 4, with a constant $K^{\ast}$ phase, are shown. The full curves represent the complete result, while for the dashed curves only the $K^{\ast}$ contribution was considered. The data are from Ref.~\cite{Boyarski69}.\label{fig: diffcs_highen}}
\end{center}
\end{figure}

\begin{figure}[ht]
\begin{center}
\includegraphics[width=0.5\textwidth]{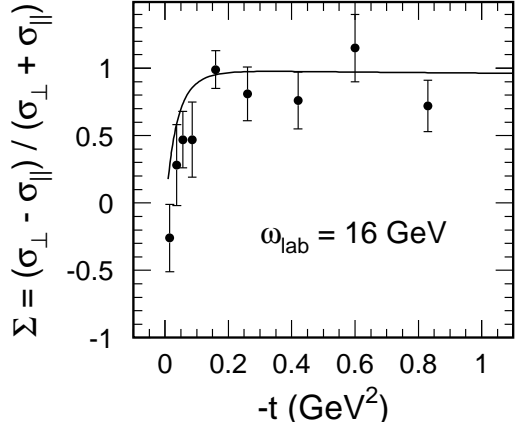}
\caption{Results for the forward-angle $p(\gamma,K^+)\Lambda$ photon beam asymmetry at $\omega_{lab}=16~\mathrm{GeV}$. The curves for the various models are virtually indistinguishable, so for the sake of clarity we display only the asymmetry for model variant 1. The data are from Ref.~\cite{Qui79}.\label{fig: phopol_highen}}
\end{center}
\end{figure}

\begin{figure}[ht]
\begin{center}
\includegraphics[width=0.68\textwidth]{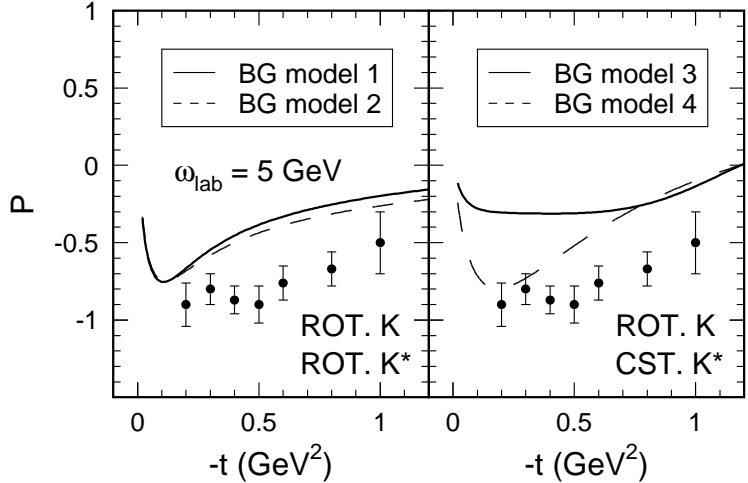}
\caption{Results for the forward-angle $p(\gamma,K^+)\Lambda$ recoil asymmetry at $\omega_{lab}=5~\mathrm{GeV}$.  The full and dashed curves in the left panel correspond to the Regge model variants 1 and 2 respectively, with a rotating phase for the $K$ and $K^{\ast}$ trajectories. In the right panel, model variants 3 (full) and 4 (dashed), with a constant $K^{\ast}$ phase, are shown. The data are from Ref.~\cite{Vo72}.\label{fig: recpol_highen}}
\end{center}
\end{figure}

Though the limited number of couplings contained in the Regge model is certainly an asset, the scarcity of the data prevents a unique determination of the $t$-channel background dynamics. We have investigated three of the four possible phase combinations for the $K$ and $K^{\ast}$ Regge propagators: rotating $K$ and $K^{\ast}$ phases, constant $K$ plus rotating $K^{\ast}$ phase, and rotating $K$ plus constant $K^{\ast}$ phase. The option with a constant phase for both propagators has not been considered, because the corresponding Regge amplitude has no imaginary part. This would result in a zero recoil polarization, contradicting experiment. Ultimately, we have identified a total of four plausible $t$-channel Regge model variants capable of describing the high-energy data in a satisfactory way. The model specifications are summarized in Table~\ref{tab: bg_specific}. 
\begin{table}[b]
\caption{Comparison of the Regge model variants (numbered 1 through 4) found to describe the high-energy, forward-angle $p(\gamma,K^+)\Lambda$ data~\cite{Boyarski69,Vo72,Qui79}. The $K$ and $K^{\ast}$ trajectory phase options are given in the second column, while the last column shows the attained $\chi^2$ value. The remaining columns contain the extracted background parameters.\label{tab: bg_specific}}
\begin{ruledtabular}
\begin{tabular}{l l  r r r r}
BG model & $K$/$K^{\ast}$ traj.~phase & $\frac{g_{K\Lambda p}}{\sqrt{4 \pi}} $ & $G^v_{K^{\ast}}$ & $G^t_{K^{\ast}}$ & $\chi^2$ \hspace{-3.2pt} \rule[-10pt]{0pt}{21pt} \\
\hline 
\rule[-0pt]{0pt}{10pt}\hspace{-3.2pt}
1 & rot.~$K$, rot.~$K^{\ast}$ & -3.23  & 0.281 & 1.09 & 3.17\\
2 & rot.~$K$, rot.~$K^{\ast}$ & -3.20  & 0.288 & -0.864 & 2.73\\
3 & rot.~$K$, cst.~$K^{\ast}$ & -3.00 & -0.189 & 1.17 & 4.37\\
4 & rot.~$K$, cst.~$K^{\ast}$ & -3.31 & -0.350 & -0.703 & 3.37\rule[-3pt]{0pt}{2pt} \\
\end{tabular}
\end{ruledtabular}
\end{table}
All model variants with a constant $K$ and a rotating $K^{\ast}$ phase resulted in unsatisfactory values of $\chi^2$, of the order of 6.5. This can be attributed to the recoil asymmetry, which was found to be the observable most discriminative with respect to the Regge model variant used. The calculated high-energy observables for each of the four model variants are compared to the data in Figs.~\ref{fig: diffcs_highen}-\ref{fig: recpol_highen}. 

The differential cross sections are displayed in Fig.~\ref{fig: diffcs_highen}.
It turns out that this observable is rather insensitive to the chosen Regge propagator phases. Indeed, for each of the two investigated phase combinations shown, as well as for the constant $K$ and rotating $K^{\ast}$ phase option, a fair description of the cross sections can be achieved for certain combinations of the parameters.

At the level of the unpolarized observables, there is only one notable difference between the model variants with two rotating phases, and the ones with a constant plus a rotating phase. While the former option results in differential cross sections that fall steadily with $t$, the latter leads to a smooth oscillatory behavior of the cross sections. This effect can be attributed to interference between the diagrams with $K$ and $K^{\ast}$ propagators. When \emph{e.g.} the $K^{\ast}$ phase is constant, the interference terms in question have a phase $e^{\pm i\pi\alpha_K(t)}$. Thus, their contribution to the differential cross section is proportional to 
\begin{equation}
\begin{split}
\mathcal{M}_{interf.}^{~K-K^{\ast}} &\sim \ \mathcal{R}e \, (\mathcal{P}_K\mathcal{P}^{~\dst\ast}_{K^{\ast}}) \ \sim \ \cos\pi\alpha_K(t) \\ 
&= \ \cos \left\{2\pi\left(\frac{t}{2.9~\mathrm{GeV}^{2}} - 0.085\right)\right\}\,,
\end{split}
\end{equation}
since in the unpolarized cross section only the real part of the propagator product $\mathcal{P}_K\mathcal{P}^{~\dst\ast}_{K^{\ast}}$ is retained. This corresponds to a harmonic oscillation in $t$, with a period of $2.9~\mathrm{GeV}^{2}$. For the situation with two rotating trajectory phases, the interference term is proportional to $\cos\pi(\alpha_K(t)-\alpha_{K^{\ast}}(t))$, which has a considerably longer oscillation period of $13.3~\mathrm{GeV}^{2}$. 

Another interesting feature of the differential cross sections is the plateau at extreme forward angles ($t\rightarrow 0$). This particular behavior cannot be reproduced in a model with only the $K$ and $K^{\ast}$ exchange diagrams from Fig.~\ref{fig:feyndiag_t} (a) and (b). This is due to the specific structure of the $\gamma KK$ and $\gamma K K^{\ast}$ Lagrangians of Eqs.~(\ref{eq:gkk}) and (\ref{eq:gkkstar}), which result in electromagnetic vertex factors going to zero at $t=0$. The presence of a Regge propagator does not alter that fact since, at low $\abs{t}$, it approaches the Feynman propagator by construction. Figure~\ref{fig: diffcs_highen} illustrates the above for the $\gamma K K^{\ast}$ interaction, by also showing the $K^{\ast}$ contribution (dashed curves) to the differential cross section. Traditionally, the issue of pure $t$-channel mechanisms being insufficient to describe the data was resolved by resorting to so-called (over)absorption mechanisms~\cite{Irv77}. 
The underlying principle is, simply stated, the following. Elastic and inelastic rescattering of the initial $\gamma p$ and final $K^+\Lambda$ states result in a loss of flux, and thus a reduction of the $p(\gamma,K^+)\Lambda$ amplitude. Hereby, the lower partial waves are absorbed most. As a consequence, the sum of all reduced partial-wave amplitudes will no longer be identically zero at $t=0$. Although this prescription is quite effective in describing the cross-section data, it results in an unphysical change of sign of the lowest partial waves. In the model presented here, the plateau in the differential cross section is naturally reproduced through the inclusion of the gauge-restoring $s$-channel electric Born term [Eq.~(\ref{eq: gauge_recipe})]. Due to its vertex structure, this diagram has an amplitude which peaks at extreme forward angles. When $\abs{t}$ increases, its influence gradually diminishes and, as is clear from Fig.~\ref{fig: diffcs_highen}, the $K^{\ast}$ exchange diagram starts dominating the process. The $K$ contribution remains quite modest in the entire $t$ region under consideration, since $\mathcal{P}_{Regge}\sim s^{\alpha(t)}$ and the $K$ trajectory has a smaller offset than the $K^{\ast}$ trajectory [see Eqs.~(\ref{eq: Ktraj}) and (\ref{eq: Kstartraj})]. The inclusion of the $s$-channel electric Born diagram can result in either a peak or a plateau, depending on the relative values of the coupling constants of the $K^{\ast}$ exchange and nucleon-pole diagram. In the $p(\gamma,K^+)\Lambda$ case, the interplay between both diagrams results in a plateau. 

Figure~\ref{fig: phopol_highen} displays the photon beam asymmetry $\Sigma$ at $\omega_{lab}=16~\mathrm{GeV}$. 
This observable is extremely well reproduced in all four models presented here. Its insensitivity to the particular choice of background model is even more pronounced than for the differential cross sections. Only the result for model variant 1 is shown, since the three other curves are nearly identical. The asymmetry is small at extreme forward angles, rising quickly towards 1. The fact that the $\sigma_{\perp}$ contribution dominates at higher $\abs{t}$ indicates that a natural-parity particle, here the $K^{\ast}$, is exchanged.  The exchange of the unnatural-parity $K$ mostly influences $\sigma_{\Vert}$, while the $s$-channel Born diagram contributes more or less equally to $\sigma_{\perp}$ and $\sigma_{\Vert}$. This explains the behavior of $\Sigma$ at forward angles, where the dynamics are mostly governed by the $s$-channel Born diagram.

To our knowledge, the sole high-energy $p(\gamma,K^+)\Lambda$ recoil asymmetry data available were collected at a photon energy of $5~\mathrm{GeV}$ in the early seventies ~\cite{Vo72}. A comparison with our results for the different Regge model variants is presented in Fig.~\ref{fig: recpol_highen}. 
Since the measured asymmetry is nonzero, we conclude that the $t$-channel dynamics are governed by the exchange of two or more trajectories. Indeed, polarized baryon asymmetries reflect interference effects, requiring at least two non-vanishing contributions to the amplitude, with different phases. 

The recoil asymmetry is an extremely useful observable for constraining the reggeized background dynamics. Firstly, since it is proportional to $\mathcal{I}m \, (\mathcal{P}_K\mathcal{P}^{~\dst\ast}_{K^{\ast}})$, the assumption of a constant $K$ phase would lead to a $\sin \pi\alpha_{K^{\ast}}(t)$ dependence for $P$. The calculated asymmetry would then be exactly zero at $t=-0.38~\mathrm{GeV}^2$. This is, however, precisely the point where the measured asymmetry reaches its maximum. This explains why the possibility of a constant $K$ trajectory phase is rejected. Secondly, the sign of the recoil asymmetry is directly linked to the relative signs of the $g_{K\Lambda p}$ and $G^{v,t}_{K^{\ast}}$ couplings. Indeed, Table~\ref{tab: bg_specific} shows that the negative sign for the recoil asymmetry imposes severe constraints upon the signs of the $K^{\ast}$ couplings. For a rotating $K^{\ast}$ phase, the coupling $G^v_{K^{\ast}}$ should be positive; for a constant $K^{\ast}$ phase a negative vector coupling is needed. The sign of the tensor coupling appears to be of less importance.

\subsection{Extrapolation to the resonance region}
\label{subsec: results res}

The latest $p(\gamma,K^+)\Lambda$ resonance-region data provided by CLAS and SAPHIR consist of differential and total cross sections and hyperon polarizations~\cite{Glander04,McNabb03,Brad05}. Photon beam asymmetries for the forward-angle kinematical region have been supplied by LEPS~\cite{Zegers03,Sumihama05}. The discrepancy between the CLAS and SAPHIR data has been heavily discussed. This issue has been largely resolved with the release of new cross-section results by the CLAS collaboration, which are in fair to good agreement with the SAPHIR ones~\cite{Brad05}. We shall limit our analysis to the results from CLAS, as they are the most recent. Since the forward-angle kinematical region constitutes the main focus of this work, we shall only consider the $\cos\theta_K^{\ast} > 0.35$ part of the CLAS data in our analysis ($\theta_K^{\ast}$ being the kaon scattering angle in the center-of-mass frame), supplemented by the LEPS photon asymmetry data from Ref.~\cite{Zegers03}.

In contrast to its smoothness at high $\omega_{lab}$, the $p(\gamma,K^+)\Lambda$ cross section exhibits a richer structure at lower energies. This hints at individual $s$-channel resonances contributing to the photoproduction dynamics. Clearly, no resonant effects arise from the Regge amplitudes. To remedy this, we superimpose a number of resonance terms onto the reggeized background, in such a way that they vanish in the high-$s$ limit. As explained in Sec.~\ref{subsec: RPR res}, the latter is accomplished by introducing a Gaussian form factor at each of the strong $K\Lambda N^{\ast}$ vertices. We deliberately keep the model uncertainties at a strict minimum. Accordingly, resonances with spin $J$ larger than 3/2 are not taken into account because the corresponding Lagrangians cannot be given in an unambiguous way. We also refrain from including resonances with a mass above 2~GeV, in order to minimize any double-counting effects that might arise from superimposing a large number of individual $s$-channel diagrams onto the $t$-channel Regge amplitude. 

Previous studies attributed a sizable role to the $S_{11}(1650)$, $P_{11}(1710)$ and $P_{13}(1720)$ states~\cite{FeMo99,MaBe00,Saghai01}. This ``core'' set of resonances generally falls short of fully reproducing the experimental results for $1.3~\mathrm{GeV} \leq \omega_{lab} \leq 1.6~\mathrm{GeV}$ (or, $1.8~\mathrm{GeV} \leq W \leq 2~\mathrm{GeV}$). The Particle Data Group mentions the two-star $P_{13}(1900)$ as the sole nucleon resonance in the 1900-MeV mass region ~\cite{PDG04}. Taking into account the width cited for this resonance, which is as high as 500 MeV, it appears unlikely that the $P_{13}(1900)$ state by itself can explain the quite narrow structure in the forward-angle cross sections. Possibly a second, as yet unknown resonance is manifesting itself in the $p(\gamma,K^+)\Lambda$ observables. Following the discussion in Refs.~\cite{Saghai01,MaSuBe04,Sara05,GApaper04}, we investigate the options of a missing $D_{13}(1900)$ or $P_{11}(1900)$ state. 

It should be noted that the much-debated $W \approx 1900$ MeV cross-section peak is angle dependent in position and shape~\cite{Brad05}, possibly hinting at the interference of two or more resonances with a mass in the indicated $W$ range. Alternatively, photoproduction of an $\eta$ particle or of $K\Lambda^{\ast}$ and $K^{\ast}\Lambda$ states could also lead to additional structure in the observables through final-state interactions. Whatever the underlying explanation, reproducing the angle-dependence of this structure will likely be easier to accomplish in a Regge-inspired model than in the standard isobar approaches, since reggeization requires the forward- and backward-angle kinematical regions to be treated separately. Here, we shall direct our efforts towards the forward-angle observables only.  An investigation of the structure appearing at backward angles would require $u$-channel reggeization. 

To our knowledge, the only other study of the $p(\gamma,K^+)\Lambda$ process carried out in a mixed Regge-isobar framework is the one by Mart and Bennhold (MB)~\cite{MaBe04}. It differs from ours in many respects. Most importantly, the MB model combines the high-energy $t$-channel Regge amplitude with the \emph{full} Feynman amplitude for the resonance region. Contrary to our RPR approach, isobar background terms are explicitly included, resulting in a larger number of free parameters. Furthermore, the transition from the resonance to the high-energy region involves a phenomenological mixing of the Regge and isobar parts. Finally, Mart and Bennhold have considered only the case of rotating phases for the $K$ and $K^{\ast}$ trajectories. With regard to the $s$-channel resonances, the MB model contains the same ``core'' set as ours, supplemented with a $D_{13}(1900)$ state.

\renewcommand{\arraystretch}{0.85}
\begin{table}[b]
\caption{Sets of $N^{\ast}$ resonances used in the calculations. Each of the six $N^{\ast}$ sets has been combined with all four background model options (1 through 4) from Table~\ref{tab: bg_specific}, resulting in a total of 24 RPR model variants (1a, 1b etc.). The last column mentions the number of free parameters. This number does not include background couplings, since they were fixed against the high-energy data. \label{tab: allRPR}}
\begin{ruledtabular}
\begin{tabular}{c  c c c c c c c}
\normalsize{$N^{\ast}$} \quad & $S_{11}(1650)$, $P_{11}(1710)$, & $P_{13}(1900)$ & $D_{13}(1900)$ & $P_{11}(1900)$ & NFP \\ 
\quad\normalsize{set}\quad\quad & $P_{13}(1720)$ (``core'') & (PDG) & (``missing'') & (``missing'') & \\ 
\hline
a & $\bigstar$ & -- & -- & -- & 8 \\
b & $\bigstar$ & $\bigstar$ & -- & -- & 13 \\ 
c & $\bigstar$ & -- & $\bigstar$ & -- & 13 \\
d & $\bigstar$ & -- & -- & $\bigstar$ & 9 \\
e & $\bigstar$ & $\bigstar$ & $\bigstar$ & -- & 18 \\ 
f & $\bigstar$ & $\bigstar$ & -- & $\bigstar$ & 14 \\
\end{tabular}
\end{ruledtabular}
\end{table}

Table~\ref{tab: allRPR} gathers the various combinations of nucleon resonances used in our calculations. Each of the six $N^{\ast}$ sets (a through f) from Table~\ref{tab: allRPR} was combined with the four background model variants (1 through 4) from Table~\ref{tab: bg_specific}, resulting in twenty-four RPR model variants to be considered. Of these twenty-four, three combinations stood out as providing the best global description of the high- and low-energy observables. They are labeled RPR-2, RPR-3 and RPR-4, corresponding to background model variants 2, 3 and 4 respectively. Their properties are summarized in Table~\ref{tab: res_specific}. The $\chi^2$ values result from a comparison of the $p(\gamma,K^+)\Lambda$ calculations with the data in the resonance and high-energy regions~\cite{McNabb03, Brad05, Zegers03, Boyarski69,Vo72,Qui79}. Apart from these ``raw'' $\chi^2$ values, the number of free parameters also serves as an important criterion by which to compare the different model variants. This number is mentioned in both Tables~\ref{tab: allRPR} and~\ref{tab: res_specific}. It is worth stressing that in our RPR approach, the only parameters that remain to be fitted to the resonance-region data are the resonance couplings and the cutoff $\Lambda_{res}$ for the strong resonance form factors. As described in the previous section, the background has been fixed against the high-energy observables. Moreover, for the masses and widths of the known resonances we have assumed the PDG values~\cite{PDG04} instead of treating them as additional free parameters as is often done. For $\Lambda_{res}$, values between 1400 and 2200 MeV were obtained. These values for the resonance cutoff are compatible with those typically used for the dipole form factors assumed in isobar models.

\begin{table}[t]
\caption{RPR model variants providing the best description of the $p(\gamma,K^+)\Lambda$ data from threshold up to a photon energy of 16 GeV. Apart from the information also contained in Table~\ref{tab: allRPR}, the Regge background model (BM) is given (using the numbering from Sec.~\ref{subsec: results background}), as are the resonance cutoff $\Lambda_{res}$ and the attained value of $\chi^2$.
\label{tab: res_specific}}
\begin{ruledtabular}
\begin{tabular}{l c c c c c c c c}
no. & BM & core & $P_{13}(1900)$ & $D_{13}(1900)$ & $P_{11}(1900)$ & $\Lambda_{res}$ (MeV) & NFP & $\chi^2$\vspace*{2mm} \\ 
\hline
RPR-2 & 2 & $\bigstar$ & $\bigstar$ & -- & $\bigstar$ & 2160 & 14 & 3.0 
\\
RPR-3 & 3 & $\bigstar$ & $\bigstar$ & -- & $\bigstar$ & 1800 & 14 & 3.1
\\
RPR-4 & 4 & $\bigstar$ & -- & -- & -- & 1405 & 8 & 3.3 \\
\end{tabular}
\end{ruledtabular}
\end{table} 

The background option 1 is missing from Table~\ref{tab: res_specific}, as it failed to produce acceptable results in combination with any of the proposed $N^{\ast}$ sets. This prompts the conclusion that the assumption of a rotating phase for both the $K$ and $K^{\ast}$ trajectories, in combination with positive $G^{v,t}_{K^{\ast}}$ couplings, is ruled out by the resonance-region data. We stress again that the high-energy observables alone \emph{do not} allow one to distinguish between the background model variants 1 through 4. In fact, the background option 1 closely resembles the Regge model originally proposed by Guidal, Laget and Vanderhaeghen for the description of high-energy electromagnetic production of kaons from the proton~\cite{Guidalthes,reg_guidal, reg_guidal_2}. 

Interestingly, the RPR model variant based on background option 4 did not require a contribution from any resonance other than those contained in the ``core''. The effect on $\chi^2$ of including the two-star $P_{13}(1900)$ turned out to be significant for the background model variants 2 and 3 only, as did the assumption of an extra, as yet unmeasured resonance. The RPR-4 model thus contains a significantly smaller number of free parameters than the two other options (8 as opposed to 14). On this basis, we are tempted to consider the RPR-4 option the ``best'' model in this study. For the background models 2 and 3, our calculations point to a $P_{11}(1900)$ state as the most likely missing-resonance candidate. The RPR model variants containing a $D_{13}$ resonance resulted in a higher value of $\chi^2$ than those with a $P_{11}$ state, despite having four free parameters more. 
Still, the difference in $\chi^2$ between the models with a $D_{13}$ and those with a $P_{11}$ is rather small, the former choice leading to values only about 0.5 higher than the latter. In other words, the presence of a $D_{13}(1900)$ state, while less likely, cannot be ruled out entirely.

\begin{figure*}[t]
\begin{center}
\includegraphics[width=0.75\textwidth]{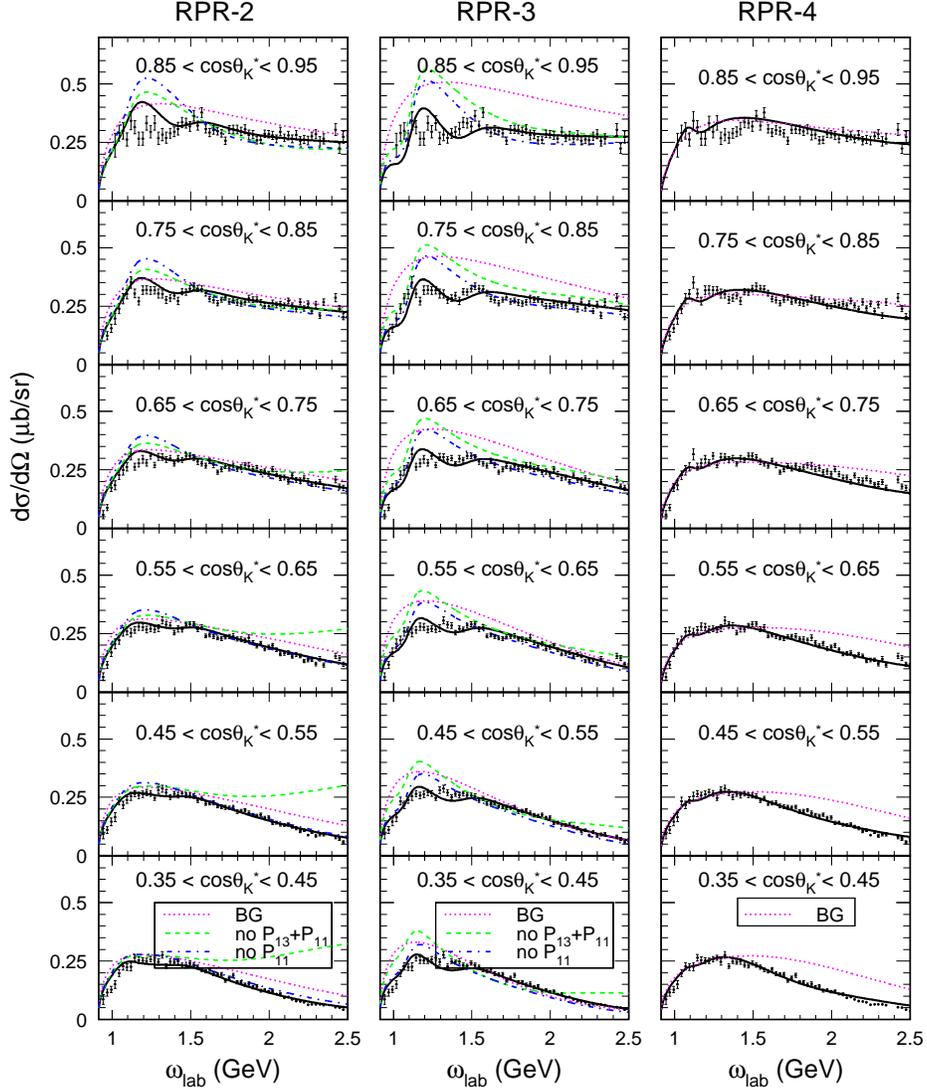}
\caption{Energy dependence of the differential $p(\gamma,K^+)\Lambda$ cross sections in the resonance region, for all bins of $\cos\theta_{K}^{\ast}$ considered in the fitting procedure. The RPR-2 (left) and RPR-3 (center) model calculations include the two-star $P_{13}(1900)$ and the missing $P_{11}(1900)$, while the the RPR-4 (right) option only contains the ``core'' resonances. The full curves represent the complete result, while the dotted, dashed and dash-dotted curves show the various contributions, as indicated in the figure. The data are from CLAS~\cite{Brad05}. 
\label{fig: diffcs_lowen} }
\end{center}
\end{figure*}

Figure~\ref{fig: diffcs_lowen} shows the energy dependence of the differential cross sections for the three RPR model variants presented in Table~\ref{tab: res_specific}. The results confirm that the Regge background in itself produces cross sections having approximately the right order of magnitude. We also note the smoothness of the Regge background as compared to the experimental cross sections. 
It is obvious from Fig.~\ref{fig: diffcs_lowen} that the missing $P_{11}(1900)$ resonance plays a vital role in the RPR-2 and RPR-3 calculations. This is reflected in the fitted resonance couplings, given in Table~\ref{tab: res_couplings}. Remarkably, even though the RPR-4 option does not contain any resonances with a mass of about 1900 MeV, it results in cross sections with a clearly visible structure around this energy. This observation corroborates that the presence of a cross-section peak does not necessarily point to a resonance, but may also be explained by tuning the background~\cite{Saghai01} or by including channel-coupling effects~\cite{Mosel02_pho,MoShklyar05}. 

Our conclusions concerning the existence of a missing $M=1900~\mathrm{MeV}$ state evidently depend on the Regge background model used. Judging by the cross-section data alone, no convincing evidence for a missing resonance is found when the RPR model 4 is assumed, while the models RPR-2 and RPR-3 do require the presence of such a state. We recall that the Regge model variants 3 and 4 differ in the sign of the $G^t_{K^{\ast}}$ coupling. This sign turned out to be of minor importance in describing the high-energy data. Apparently, the impact of the tensor interaction in the strong $K^{\ast} \Lambda p$ vertex [see Eq.~(\ref{eq:VLP})] grows as lower energies are probed.

Finally, since the energy behavior of the differential cross sections is well reproduced in all three of the proposed RPR models, we may conclude that the somewhat problematic ``flattening out'' of the resonance peaks as observed by Mart and Bennhold in their calculated cross-sections~\cite{MaBe04} is not an issue when the Regge and isobar approaches are combined via the RPR prescription.

\renewcommand{\arraystretch}{1.1}
\begin{table}[t]
\begin{center}
\caption{Resonance parameters for the three proposed RPR model variants. The parameters for the ``core'' resonances are given in the upper table. The lower table mentions the couplings for the $P_{13}(1900)$ and $P_{11}(1900)$ resonances, included in the RPR-2 and RPR-3 models only.  \label{tab: res_couplings}}
\begin{tabular*}{\textwidth}{|c | c | @{\extracolsep{\fill}}c | c|c|c|c|c|}
\hline
\textbf{~RPR~} & \,$S_{11}(1650)$\, & \,$P_{11}(1710)$\, & \multicolumn{5}{c|}{$P_{13}(1720)$} \\ \cline{2-8}
\textbf{mod.} & $G_{S_{11}(1650)}$ & $G_{P_{11}(1710)}$ & $G^{(1)}_{P_{13}(1720)}$ & $G^{(2)}_{P_{13}(1720)}$ & $X_{P_{13}}$ & $Y_{P_{13}}$ & $Z_{P_{13}}$ \\
\hline\hline
2 & $-3.34 \cdot 10^{-2}$ & $-1.12 \cdot 10^{-1}$ & $-1.21 \cdot 10^{-2}$ & $-4.28 \cdot 10^{-3}$ & $-2.3 \cdot 10^{2}$ & $-1.65 \cdot 10^1$ & $-1.41$ \\
\hline
3 & $-1.12 \cdot 10^{-3}$ & $-3.15 \cdot 10^{-1}$ & $1.54 \cdot 10^{-4}$ & $1.34 \cdot 10^{-2}$ & $-3.40$ & $1.21 \cdot 10^2$ & $-5.08$ \\
\hline
4 & $1.22 \cdot 10^{-3}$ & $2.35\cdot 10^{-1}$ & $-1.36 \cdot 10^{-4}$ & $4.72 \cdot 10^{-1}$ & $-1.12 \cdot 10^1$ & $3.14 \cdot 10^{3}$ & $-9.64 \cdot 10^{-1}$ \\
\hline
\end{tabular*}

\vspace*{4mm}

\begin{tabular*}{\textwidth}{| c| c|@{\extracolsep{\fill}}c|c|c|c |c|}
\hline
\textbf{~RPR~} & \multicolumn{5}{c|}{$P_{13}(1900)$} & \,$P_{11}(1900)$\, \\ 
\cline{2-7}
\textbf{mod.} & $G^{(1)}_{P_{13}(1900)}$ & $G^{(2)}_{P_{13}(1900)}$ & $X_{P_{13}}$ & $Y_{P_{13}}$ & $Z_{P_{13}}$ & $G_{P_{11}(1900)}$ \\
\hline\hline
2 & $-4.96 \cdot 10^{-2}$ & $-1.41 \cdot 10^{-2}$ & $4.18 \cdot 10^{1}$ & $3.58$ & $-7.60 \cdot 10^{-1}$ & $-2.22 \cdot 10^{-1}$ \\
\hline
3 & $6.50 \cdot 10^{-2}$ & $2.70 \cdot 10^{-1}$ & $-2.07 \cdot 10^{1}$ & $2.59 \cdot 10^1$ & $-1.91 \cdot 10^{-1}$ & $-2.66 \cdot 10^{-1}$ \\
\hline
\end{tabular*}
\end{center}
\end{table} 

\begin{figure*}[ht]
\begin{center}
\includegraphics[width=0.78\textwidth, clip]{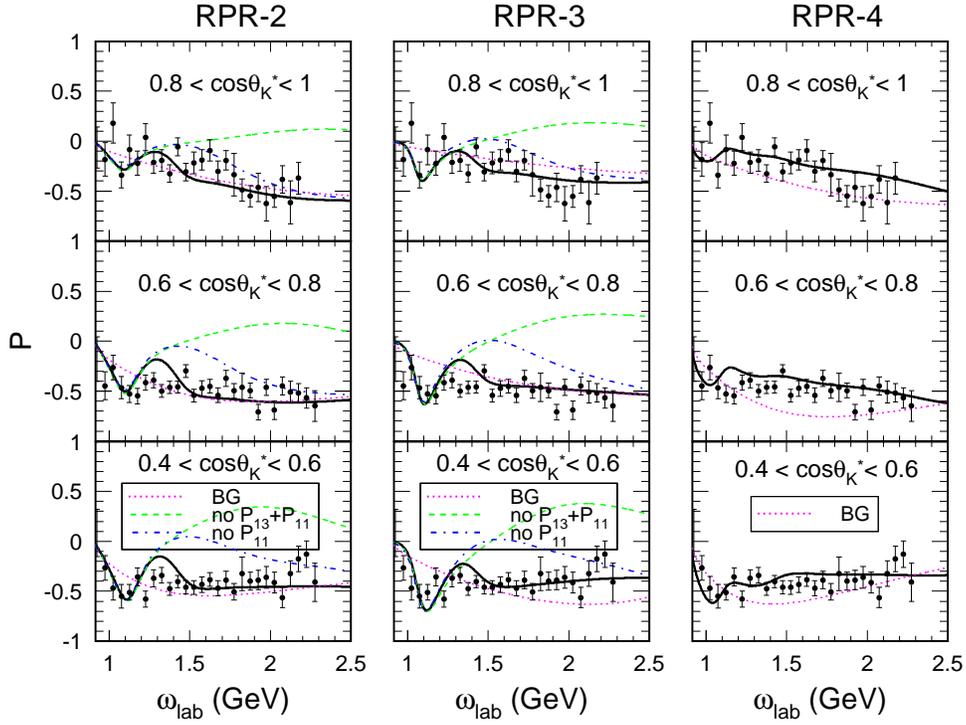}
\caption{Energy dependence of the $p(\gamma,K^+)\Lambda$ recoil polarization for those bins of $\cos\theta_{K}^{\ast}$ considered in the fitting procedure. The RPR-2 (left) and RPR-3 (center) model calculations include the two-star $P_{13}(1900)$ and the missing $P_{11}(1900)$, while the RPR-4 (right) option only contains the ``core'' resonances. Line conventions are as in Fig.~\ref{fig: diffcs_lowen}. The data are from CLAS~\cite{McNabb03}. \label{fig: recpol_lowen}}
\end{center}
\end{figure*}

The computed recoil polarizations $P$ are shown in Fig.~\ref{fig: recpol_lowen} as a function of energy.
All three RPR model calculations do an excellent job in describing this observable over the entire angular region considered for the fitting procedure ($0.35 \le \cos\theta_K^{\ast} \le 1 $). As with the unpolarized cross sections, the resonance-region data for $P$ are matched quite well by the background contribution. Interestingly, in the RPR-2 and RPR-3 models this agreement worsens once the ``core'' resonances are included. The contribution from the two-star $P_{13}(1900)$ resonance serves to lower the asymmetries to the required negative values. This does not suffice, though. An additional $P_{11}(1900)$ resonance is needed to temper the residual $\omega_{lab} \approx 1.5$ GeV ($W \approx 1.9$ GeV) peak by interfering destructively with the background and core diagrams. By contrast, in the RPR model 4, the core resonances by themselves suffice to provide the quite modest correction to the background necessary to reproduce the data. The difference between the RPR-2 and RPR-3 model variants is practically negligible as far as the recoil polarizations are concerned. In both cases, the dominant contribution to the asymmetry comes from the $P_{13}(1900)$ resonance. The influence of the $P_{11}(1900)$ manifests itself most clearly at the more backward angles considered here. The recoil polarization results thus confirm the conclusions drawn from the differential cross sections: the $P_{13}(1900)$ and $P_{11}(1900)$ states are essential when adopting background option 2 or 3, but the background model 4 does not require it.

In the model of Mart and Bennhold, a satisfactory description for the recoil polarization could not be achieved. Particularly, the dip in $P$ at $W \approx 1.75~\mathrm{GeV}$ ($\omega_{lab}\approx 1.16~\mathrm{GeV}$) for $\cos\theta_{K}^{\ast}<0.6$ proved problematic. This discrepancy was tentatively attributed to an unidentified resonance. Our RPR calculations, however, indicate otherwise. Indeed, as becomes clear from the lower graphs in Fig.~\ref{fig: recpol_lowen}, even with just the core set of three resonances all three RPR model variants reproduce the dip in the asymmetries.

\begin{figure}[b]
\begin{center}
\includegraphics[width=0.84\textwidth, clip]{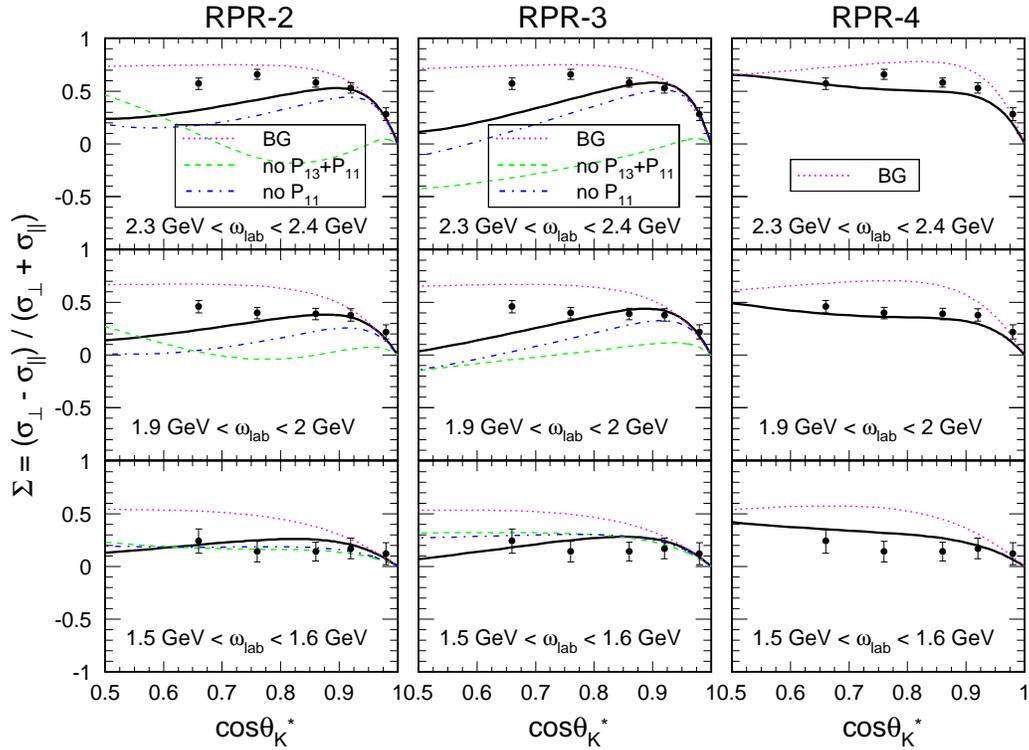}
\caption{Results for the forward-angle $p(\gamma,K^+)\Lambda$ photon beam asymmetry, for $0.5<\cos\theta_{K}^{\ast}<1.0$ and for three representative bins of $\omega_{lab}$, corresponding to center-of-mass energy bins $ 2.28~\mathrm{GeV}<W<2.32~\mathrm{GeV}$, $2.11~\mathrm{GeV}<W<2.15~\mathrm{GeV}$, and $1.92~\mathrm{GeV}<W<1.97~\mathrm{GeV}$. The RPR-2 (left) and RPR-3 (center) model calculations include the two-star $P_{13}(1900)$ and the missing $P_{11}(1900)$, while the the RPR-4 (right) option only contains the ``core'' resonances. Line conventions are as in Fig.~\ref{fig: diffcs_lowen}. The data are from LEPS~\cite{Zegers03}. \label{fig: phopol_lowen}}
\end{center}
\end{figure}

The results for the photon beam asymmetry $\Sigma$ are contained in Fig.~\ref{fig: phopol_lowen}. 
Since only a limited number of data points are available, these results might be considered more as a prediction than as the actual outcome of a fit. Each of the proposed RPR model variants reproduces this observable well at extreme forward angles. The more backward angles turn out to represent a greater challenge, however. This observation is compatible with Mart and Bennhold's findings, except that their results underestimated the low-energy data, while we find a discrepancy at the highest energies. For the photon asymmetries, the MB approach provides a better description than the model variants presented in this work. It should be realized, though, that the number of free parameters contained in the RPR model is considerably smaller. 

Summarizing, our calculations provide only circumstantial evidence for the existence of a new resonance. Although all three proposed RPR model variants result in a comparable $\chi^2$, we prefer the option based on background model 4. Indeed, it contains a significantly smaller number of free parameters than the RPR-2 and RPR-3 options (8 as opposed to 14). For the  RPR-2 and RPR-3 model variants, the inclusion of a yet unmeasured state with a mass of $1900~\mathrm{MeV}$ significantly improves the agreement between the calculations and the data. Judging from our results, the $P_{11}(1900)$ is the most likely candidate.

\section{Conclusions}
\label{sec: conclusion}

We have presented a relatively simple and economical framework for describing $p(\gamma,K^+)\Lambda$ processes from threshold up to $\omega_{lab}=16~\mathrm{GeV}$. To this end, we resorted to a ``Regge-plus-resonance'' (RPR) strategy, involving the superposition of a limited number of $s$-channel resonances onto a reggeized $t$-channel background. The Regge-inspired approach guarantees an appropriate high-energy limit, while the $s$-channel terms provide the resonant structure to the low-energy observables. 

The parameters of the $t$-channel Regge background amplitude were determined against the high-energy data ($5~\mathrm{GeV} \leq \omega_{lab} \leq 16~\mathrm{GeV}$). We addressed the question of whether a constant or a rotating phase represents the optimum choice for the $K$ and $K^{\ast}$ Regge trajectories, and identified the recoil asymmetry as the observable most sensitive to the specific ingredients of the Regge model. In particular, the option of a constant $K$ trajectory phase could be ruled out. Despite the parameter-poorness of the Regge amplitude, singling out one particular background model turned out to pose some difficulties due to the scarcity of high-energy data. Instead, several plausible options for modeling the high-energy $t$-channel amplitude had to be retained.

We added $s$-channel diagrams to the reggeized background amplitude. In order to minimize any double-counting effects that might arise, the number of resonances was deliberately constrained. Apart from the ``core'' set consisting of the $S_{11}(1650)$, $P_{11}(1710)$ and $P_{13}(1720)$ states, we investigated possible contributions of the two-star $P_{13}(1900)$ state, as well as the as yet unobserved $D_{13}(1900)$ or $P_{11}(1900)$ resonances. 

The option of rotating $K$ and $K^{\ast}$ trajectory phases, combined with positive $K^{\ast}$ vector and tensor couplings, turned out to be incompatible with the resonance-region data. It is remarkable that precisely this option has always been regarded as the ``standard'' choice for the Regge description of the high-energy observables. 

In some cases the inclusion of a new resonance, in combination with the two-star $P_{13}(1900)$, improved the agreement with the data significantly. The $P_{11}(1900)$ state emerges from our calculations as a more likely missing-resonance candidate than the $D_{13}(1900)$. Still, we are reluctant to claim evidence for the existence of either of these states. Indeed, we have shown that an equally good description of the $p(\gamma,K^+)\Lambda$ dynamics in the entire energy region under study can be achieved with a model containing only the ``core'' set of known resonances. This demonstrates that the much-discussed structure in the observables around $W \approx 1900$ MeV is not necessarily an indication of a resonance in this mass region, but may also be explained by tuning the background.

\begin{acknowledgments}
We are obliged to R. Schumacher for providing us with the latest CLAS data. We acknowledge enlightening discussions with M. Vanderhaeghen, and thank D. Ireland for a critical reading of the manuscript. This work was supported by the Fund for Scientific Research, Flanders (FWO) under contract number G.0020.03. 
\end{acknowledgments}

\appendix* 

\section{Effective-Lagrangian formalism}
\label{app: eff lagr}

\subsection{Forward-angle background}
\label{app: eff lagr background}

\noindent \emph{Electromagnetic couplings.}
The electromagnetic interaction Lagrangians contributing to the $t$-channel background amplitude are given by: 
\begin{align}
{\cal L}_{\gamma KK} &= - i e \left( K^\dagger \partial_\mu
K - K \partial_\mu K^\dagger \right) A^\mu \label{eq:gkk} \;, \\[4pt]
{\cal L}_{\gamma KK^{\ast}} &= \ \ \frac{ e \kappa_{
KK^{\ast}}}{4M} \ \epsilon^{\mu \nu \lambda \sigma} F_{\mu \nu}
V_{\lambda \sigma} K + h.\,c. \;,
\label{eq:gkkstar} \\
{\cal L}_{\gamma pp}^{elec} &= - e \ \overline{N} \gamma_\mu N A^\mu
\label{eq:gpp} \;.
\end{align}
The antisymmetric tensor for the photon field $A^{\mu}$ is defined as $F^{\mu \nu} = \partial^\nu A^\mu - \partial^\mu A^\nu$. Analogously, the vector meson tensor is given by $V^{\mu \nu} = \partial^\nu V^\mu - \partial^\mu V^\nu$. For the proton anomalous magnetic moment, $\kappa_p = 1.793~\mu_N$~\cite{PDG04} is used. By convention, the mass scale $M$ is taken at $1~\mathrm{GeV}$, and $e=+\sqrt{4\pi/137}$.\\

\noindent\emph{Strong couplings.}
For the strong $K\Lambda p$ vertex, either a pseudoscalar or a pseudovector structure can be assumed. We have opted for a pseudoscalar interaction:
\begin{equation}
{\cal L}^{PS}_{K \Lambda p} = -i g_{K\Lambda p} \ K^\dagger
\overline{\Lambda} \gamma_5 N  + h.\,c.\,, 
\label{eq: PS_KLp}\\
\end{equation}
with ``h.\,c.'' denoting the hermitian conjugate. The hadronic $K^{\ast}\Lambda p$ vertex is composed of a vector ($v$) and a tensor ($t$) part,
\begin{equation}
\begin{split}
{\cal L}_{K^{\ast} \Lambda p} =& - g^v_{K^{\ast}\Lambda p} \
\overline{\Lambda} \gamma_\mu N V^\mu  \\
&+ \frac{ g^t_{K^{\ast} \Lambda p}}{2 \left(M_{
\Lambda}+M_p \right)} \ \overline{\Lambda} \sigma_{\mu \nu} V^{\mu \nu} N + h.c.\,,
\label{eq:VLP}
\end{split}
\end{equation}
with $\sigma_{\mu\nu} = \frac{i}{2} [\gamma_{\mu}, \gamma_{\nu}]$. $V^{\mu}$ again stands for the $K^{\ast}$ vector field, and $V^{\mu\nu}$ for the corresponding antisymmetric tensor.

\subsection{Resonance contributions}
\label{app: eff lagr res}

\noindent\emph{Electromagnetic couplings.} 
The electromagnetic interaction Lagrangian for spin-$1/2$ resonances reads:
\begin{equation}
\mathcal{L}_{\gamma p N^{\ast}(\frac{1}{2})} = \frac{e \kappa_{\ssst p N^{\ast}}}{4 M_p} \
\overline{R}\,\Gamma_{\mu \nu} N F^{\mu\nu}+ h.c.\,.
\label{eq:gpr_1_2}
\end{equation}
In this expression, $R$ represents the Dirac spinor field of the resonance, and $\Gamma^{\mu\nu}=~\sigma^{\mu \nu}$ $(\gamma^5 \sigma^{\mu \nu})$ for even (odd) parity resonances.

For spin-3/2 resonances, two terms appear in the Lagrangian:
\begin{equation}
\begin{split}
\mathcal{L}_{\gamma p}&_{N^{\ast}(\frac{3}{2})} =~i \ \frac {e\kappa_{\ssst pN^{\ast}}^{\left( 1
\right) }}{2M_p}\ 
\overline{R}^\mu \theta_{\mu \nu} \left(Y \right) 
\Gamma_\lambda N F^{\lambda \nu} \\ 
&- \frac{e \kappa_{\ssst pN^{\ast}}^{\left( 2
\right)} }{4M_p^2}\ \overline{R}^\mu \theta_{\mu \nu} \left( X \right)
\Gamma \left( \partial_\lambda N \right) F^{\nu \lambda} + h.c.  
\label{eq:gpr_3_2} 
\end{split}
\end{equation}
Herein, $\Gamma = \gamma^5$ $(1)$ and $\Gamma^{\mu} = \gamma^{\mu}\gamma^5$ $(\gamma^{\mu})$ for even (odd) parity resonances. $R^{\mu}$ is the Rarita-Schwinger vector field used to describe the spin-$3/2$ particle. The function $\theta_{\mu\nu} \left(V\right)$ is defined as \cite{BeDav89}
\begin{equation}
\theta_{\mu\nu} \left(V\right) = g_{\mu\nu}-\left( V+ \frac{1}{2}
\right) \gamma_\mu \gamma_\nu \;,
\end{equation}
with $V = X,Y,Z$ the so-called \emph{off-shell parameters}. \\

\noindent\emph{Strong couplings.} 
The strong interaction Lagrangian for a spin-$1/2$ resonance can have a pseudoscalar or a pseudovector form. As in Eq.~(\ref{eq: PS_KLp}), we have used the pseudoscalar scheme:
\begin{equation}
\mathcal{L}_{K \Lambda N^{\ast}(\frac{1}{2})}^{PS} = - i g_{\ssst K \Lambda N^{\ast}} K^\dagger
\overline{\Lambda}\, \Gamma R + h.c. 
\label{eq:kyp_pv}
\end{equation}
For spin-1/2 resonance exchange, the information regarding the
extracted coupling constant takes on the form:
\begin{equation}
G_{N^{\ast}} =  \frac{g_{\ssst K\Lambda N^{\ast}}}{\sqrt{4 \pi}} \ \kappa_{\ssst pN^{\ast}} \;.
\end{equation}
The hadronic vertex for spin-$3/2$ exchange is given by
\begin{equation}
\mathcal{L}_{K \Lambda N^{\ast}(\frac{3}{2})} = \frac{f_{\ssst K \Lambda N^{\ast}}}{M_{\ssst K}}
\overline{R}^\mu \theta_{\mu\nu} \left( Z \right) \Gamma' \Lambda \left 
( \partial^\nu K \right) + h.c.\,,
\end{equation}
with $\Gamma' = 1$ $(\gamma^5)$ for even (odd) parity resonances. 

For spin-3/2 resonance exchange, the fits of the model calculations to
the data give access to the following combinations of coupling constants:
\begin{equation}
G_{N^{\ast}}^{\left( 1,2 \right)} = \frac{e f_{\ssst K\Lambda N^{\ast}}}{4\pi}
\ \kappa^{\left(1,2 \right)}_{\ssst  NN^{\ast}} \;.
\end{equation}

The normalization conventions for the field operators and Dirac matrices are those of Ref.~\cite{Peskin}. The Mandelstam variables for a two-particle scattering process of the form ${1+2\rightarrow 3+4}$ are defined in the standard way as
\begin{align}
s &= (p_1+p_2)^2\,, \\
t &= (p_1-p_3)^2\,, \\
u &= (p_1-p_4)^2\,.
\end{align}


\begin{thebibliography}{99}

\bibitem{McNabb03} J.W.C. McNabb \emph{et al.} (CLAS Collaboration), Phys. Rev. C \textbf{69}, 042201 (2004). 

\bibitem{Brad05} R. Bradford \emph{et al.} (CLAS Collaboration), nucl-ex/0509033; R. Schumacher, private communications.

\bibitem{Glander04} K.-H. Glander \emph{et al.} (SAPHIR Collaboration), Eur. Phys. J. A \textbf{19}, 251 (2004). 

\bibitem{Zegers03} R.G.T. Zegers \emph{et al.} (LEPS Collaboration), Phys. Rev. Lett. \textbf{91}, 092001 (2003). 

\bibitem{Sumihama05}  M. Sumihama \emph{et al.} (LEPS Collaboration), Nucl. Phys. \textbf{A754}, 303 (2005). 

\bibitem{MaSuBe04} T. Mart, S. Sulaksono, and C. Bennhold, in \emph{Proceedings of the International Symposium on Electrophotoproduction of Strangeness on Nucleons and Nuclei, Sendai, Japan, 2003}, edited by K. Maeda (River Edge, World Scientific, 2004). 

\bibitem{MoShklyar05} V. Shklyar, H. Lenske, and U. Mosel, Phys. Rev. C \textbf{72}, 015210 (2005). 

\bibitem{Diaz05} B. Juli\'a-Di\'az, B. Saghai, F. Tabakin, W.-T. Chiang, T.-S.H. Lee, and Z. Li, Nucl. Phys. \textbf{A755}, 463 (2005). 

\bibitem{Sara05} A.V. Sarantsev, V.A. Nikonov, A.V. Anisovich, E. Klempt, and U. Thoma, hep-ex/0506011 (2005). 

\bibitem{Stijnprc01} S. Janssen, J. Ryckebusch, D. Debruyne, and T. Van Cauteren, Phys. Rev. C \textbf{65}, 015201 (2002). 

\bibitem{AdWr88} R.A. Adelseck and L.E. Wright, Phys. Rev. C \textbf{38}, 1965 (1988). 

\bibitem{WJ92} R.A. Williams, C.-R. Ji, and S.R. Cotanch, Phys. Rev. C \textbf{46}, 1617 (1992). 

\bibitem{DaSa96} J.C. David, C. Fayard, G.H. Lamot, and B. Saghai, Phys. Rev. C \textbf{53}, 2613 (1996). 

\bibitem{MaBeHy95} T. Mart, C. Bennhold, and C.E. Hyde-Wright, Phys. Rev. C \textbf{51}, R1074 (1995). 

\bibitem{MaBe00}  T. Mart and C. Bennhold, Phys. Rev. C \textbf{61}, 012201(R) (1999). 

\bibitem{Chiang01} W.-T. Chiang, F. Tabakin, T.-S.H. Lee, and B. Saghai, Phys. Lett. B \textbf{517}, 101 (2001). 

\bibitem{Mosel02_pho} G. Penner and U. Mosel, Phys. Rev. C \textbf{66}, 055212 (2002). 

\bibitem{US05} A. Usov and O. Scholten,  Phys. Rev. C \textbf{72}, 025205 (2005). 

\bibitem{gauge_inv} K. Ohta, Phys. Rev. C \textbf{40}, 1335 (1989); H. Haberzettl, \emph{ibid.} \textbf{56}, 2041 (1997); R.M. Davidson and R. Workman, \emph{ibid.} \textbf{63}, 025210 (2001). 

\bibitem{GApaper03} S. Janssen, D.G. Ireland, and J. Ryckebusch, Phys. Lett. B \textbf{562}, 51 (2003). 

\bibitem{reg_guidal} M. Guidal, J.-M. Laget, and M. Vanderhaeghen, Nucl. Phys. \textbf{A627}, 645 (1997). 

\bibitem{reg_guidal_2} M. Vanderhaeghen, M. Guidal, and J.-M. Laget, Phys. Rev. C \textbf{57}, 1454 (1998).

\bibitem{Levy73} N. Levy, W. Majerotto, and B.J. Read, Nucl. Phys. B \textbf{55}, 493 (1973). 

\bibitem{Guidalthes} M. Guidal, Ph. D. thesis, University of Orsay (1997). 


\bibitem{Sibi03} A. Sibirtsev, K. Tsushima, and S. Krewald, Phys. Rev. C \textbf{67}, 055201 (2003). 

\bibitem{Holvoet_phd} H. Holvoet, Ph. D. thesis, Ghent University (2002). 

\bibitem{Chi03} W.-T. Chiang, S.N. Yang, L. Tiator, M. Vanderhaeghen, and D. Drechsel,  Phys. Rev. C \textbf{68}, 045202 (2003) 

\bibitem{Regge59} T. Regge, Nuovo Cim. \textbf{14}, 951 (1959). 

\bibitem{Froiss_61} M. Froissart, Phys. Rev. \textbf{123}, 1053 (1961). 

\bibitem{PDG04} S. Eidelman \emph{et al.}, Phys. Lett. B \textbf{592}, 1 (2004). 

\bibitem{Donnachie02} S. Donnachie, G. Dosch, P. Landshoff, and O. Nachtmann, \emph{Pomeron physics and QCD} (Cambridge University Press, Cambridge, 2002). 

\bibitem{Collins77} P.D.B. Collins, \emph{An introduction to Regge theory and high-energy physics} (Cambridge University Press, Cambridge, 1977). 

\bibitem{Stijnthes} S. Janssen, Ph. D. thesis, Ghent University (2002). 

\bibitem{Guidal_elec} M. Guidal, J.-M. Laget, and M. Vanderhaeghen, Phys. Rev. C \textbf{68}, 058201(R) (2003). 

\bibitem{Saghai01} B. Saghai, AIP Conf. Proc. \textbf{59}, 57 (2001), nucl-th/0105001. 

\bibitem{FeMo99} T. Feuster and U. Mosel, Phys. Rev. C \textbf{59}, 460 (1999). 

\bibitem{Tran98} M.Q. Tran \emph{et al.} (SAPHIR Collaboration), Phys. Lett. B \textbf{445}, 20 (1998). 

\bibitem{CapRob94} S. Capstick and W. Roberts, Phys. Rev. D \textbf{49}, 4570 (1994); \textbf{58}, 074011 (1998). 

\bibitem{GApaper04} D.G. Ireland, S. Janssen, and J. Ryckebusch, Nucl. Phys. \textbf{A740}, 147 (2004). 

\bibitem{Boyarski69} A.M. Boyarski \emph{et al.}, Phys. Rev. Lett. \textbf{22}, 1131 (1969). 

\bibitem{Qui79} D.J. Quinn, J.P. Rutherfoord, M.A. Shupe, D.J. Sherden, R.H. Siemann, and C.K. Sinclair, Phys. Rev. D \textbf{20}, 1553 (1979). 

\bibitem{Vo72} G. Vogel, H. Burfeindt, G. Buschhorn, P. Heide, U. K\"otz, K.-H. Mess, P. Schm\"user, B. Sonne, and B. H. Wiik, Phys. Lett. B \textbf{40}, 513 (1972). 

\bibitem{deSwart} J.J. de Swart, Rev. Mod. Phys. \textbf{35}, 916 (1963). 

\bibitem{Mac87} R. Machleidt, K. Holinde, and C. Elster, Phys. Rep. \textbf{149}, 1 (1987). 

\bibitem{Irv77} A.C. Irving and R.P. Worden, Phys. Rep. \textbf{34}, 117 (1977). 
\bibitem{MaBe04} T. Mart and and C. Bennhold, nucl-th/0412097 (2004). 


\bibitem{BeDav89} M. Benmerrouche, R.M. Davidson, and N.C. Mukhopadhyay, Phys. Rev. C \textbf{39}, 2339 (1989). 

\bibitem{Peskin} M.E. Peskin and D.V. Schroeder, \emph{An Introduction to Quantum Field Theory} (Perseus Books, Reading, Massachusetts, 1995). 

\end{thebibliography}
\end{document}